\documentclass{article}

\usepackage[final]{neurips_2025}

\usepackage[utf8]{inputenc} 
\usepackage[T1]{fontenc}    
\usepackage{hyperref}       
\usepackage{url}            
\usepackage{booktabs}       
\usepackage{amsfonts}       
\usepackage{nicefrac}       
\usepackage{microtype}      
\usepackage{xcolor}         
\usepackage{colortbl}
\usepackage{xspace}
\usepackage{graphicx} 
\usepackage{pifont}
\usepackage{amsmath}
\usepackage{tikz}
\usepackage[most]{tcolorbox}

\usepackage{multirow}
\usepackage{array}
\usepackage{booktabs}

\definecolor{blueboxcolor}{HTML}{1E90FF}
\newtcolorbox{theobox}{
  colback=blueboxcolor!10,  
  colframe=blueboxcolor!10, 
  coltitle=black,        
  fonttitle=\bfseries,   
  boxrule=0.5pt,         
  arc=0pt,               
  left=3pt, right=3pt, top=3pt, bottom=3pt 
}

\tcbset{
  mydefault/.style={
    colback=white,    
    colframe=gray!30, 
    coltitle=black,
    boxrule=0.3pt,
    arc=1mm,                   
    fonttitle=\bfseries\small,   
    enhanced jigsaw,             
    sharp corners             
  },
  every box/.style={mydefault}
}

\definecolor{redboxcolor}{HTML}{CD5C5C}
\newtcolorbox{redtheobox}{
  colback=redboxcolor!10,  
  colframe=redboxcolor!10, 
  coltitle=black,        
  fonttitle=\bfseries,   
  boxrule=0.5pt,         
  arc=0pt,               
  left=3pt, right=3pt, top=3pt, bottom=3pt 
}

\newcommand{\lightmidrule}{\arrayrulecolor{gray!50}\midrule\arrayrulecolor{black}}
\definecolor{airforceblue}{rgb}{0.36, 0.54, 0.66}
\definecolor{azure(colorwheel)}{rgb}{0.0, 0.5, 1.0}

\definecolor{yellowcolorone}{HTML}{f0eac9}
\definecolor{yellowcolortwo}{HTML}{e5e5da}
\definecolor{purplecolor}{HTML}{d1d5e6}

\newenvironment{changemargin}[2]{\begin{list}{}{
	\setlength{\topsep}{0pt}\setlength{\leftmargin}{0pt}
	\setlength{\rightmargin}{0pt}
	\setlength{\listparindent}{\parindent}
	\setlength{\itemindent}{\parindent}
	\setlength{\parsep}{0pt plus 1pt}
	\addtolength{\leftmargin}{#1}\addtolength{\rightmargin}{#2}
	}\item}
	{\end{list}}

\newcommand{\system}{\textsc{CyberTeam}\xspace}
\newcommand{\nfunc}{9\xspace}
\newcommand{\ntask}{30\xspace}
\newcommand{\ndb}{23\xspace}

\newcommand{\hlcell}{\cellcolor{blueboxcolor!12}}

\title{Benchmarking LLMs in an Embodied Environment for Blue Team Threat Hunting}

\usepackage{authblk}
\author[1]{Xiaoqun Liu}
\author[2]{Feiyang Yu}
\author[3]{Xi Li}
\author[4]{Guanhua Yan}
\author[4]{Ping Yang}
\author[4]{Zhaohan Xi}

\affil[1]{Southern University of Science and Technology}
\affil[2]{Duke University}
\affil[3]{University of Alabama at Birmingham}
\affil[4]{Binghamton University}

\affil[ ]{%
    \texttt{\textsuperscript{1}liuxq2021@mail.sustech.edu.cn}, 
    \texttt{\textsuperscript{2}fy66@duke.edu}, 
    \texttt{\textsuperscript{3}XiLiUAB@uab.edu}, \\
    \texttt{\textsuperscript{4}\{ghyan, pyang, zxi1\}@binghamton.edu}
}
\begin{document}

\maketitle

\begin{abstract}
    As cyber threats continue to grow in scale and sophistication, blue team defenders increasingly require advanced tools to proactively detect and mitigate risks. Large Language Models (LLMs) offer promising capabilities for enhancing threat analysis. However, their effectiveness in real-world blue team threat-hunting scenarios remains insufficiently explored. In this paper, we present \system, a benchmark designed to guide LLMs in blue teaming practice. \system constructs an embodied environment in two stages. First, it models realistic threat-hunting workflows by capturing the dependencies among analytical tasks from threat attribution to incident response. Next, each task is addressed through a set of embodied functions tailored to its specific analytical requirements. This transforms the overall threat-hunting process into a structured sequence of function-driven operations, where each node represents a discrete function and edges define the execution order. Guided by this framework, LLMs are directed to perform threat-hunting tasks through modular steps. Overall, \system integrates \ntask tasks and \nfunc embodied functions, guiding LLMs through pipelined threat analysis. We evaluate leading LLMs and state-of-the-art cybersecurity agents, comparing \system's embodied function-calling against fundamental elicitation strategies. Our results offer valuable insights into the current capabilities and limitations of LLMs in threat hunting, laying the foundation for the practical adoption in real-world cybersecurity applications.
\end{abstract}
\section{Introduction}

The increasing frequency and sophistication of cyber threats continue to pose significant challenges to organizational security. In 2024 alone, over $11,000$ more ($38\%$ increase!) vulnerabilities were reported compared to 2023, as evidenced by the MITRE CVE database \cite{mitre_cve}. Defenders, commonly known as the {\bf blue team}  \cite{diogenes2018cybersecurity, rajendran2011blue}, are under increasing pressure to identify, analyze, and respond to malicious activities in a timely and accurate manner, a process termed as {\bf threat hunting}. Traditionally, threat hunting has been a labor-intensive process, relying heavily on the expertise of analysts to sift through logs, correlate indicators of compromise (IOCs), and construct hypotheses about potential attacks \cite{dash2022review,gioti2024advancements}. This process demands both deep domain knowledge and an ability to integrate fragmented evidence from multiple sources under time constraints~\cite{alshumrani2023unified,yu2010vqsvm,zhou2017knowledge}.

Recent advances in Large Language Models (LLMs) have demonstrated impressive potential to augment cybersecurity practices, including malware analysis \cite{abusitta2021malware, al2024exploring, qian2025lamd, devadiga2023gleam}, penetration testing \cite{deng2023pentestgpt, deng2024pentestgpt, happe2023getting, muzsai2024hacksynth}, and fuzzing \cite{zhang2025llms, oliinyk2024fuzzing, black2024evaluating}. Building on this progress, there is growing interest in leveraging LLMs to automate or assist in threat hunting, enabling blue team defenders to scale their investigations across complex threat landscapes and respond to incidents more effectively. However, despite this momentum, the application of LLMs in blue team threat hunting remains underdeveloped. Existing frameworks tend to focus on isolated analytical tasks \cite{sehgal2023cybersecurity, faghihi2023gluecons, dash2022review, borghesi2020improving, groza2015capturing}, such as generating advisory recommendations without integrating earlier steps like threat group attribution. This fragmented design limits our understanding of how LLMs perform within complex, interdependent threat-hunting workflows.

To address this gap, we introduce \system, a practical benchmark designed to rigorously evaluate and guide the use of LLMs in blue team threat hunting. \system supports blue-team threat-hunting workflows through the following aspects:

\begin{figure}[!tp]
    \centering
    \includegraphics[width=\textwidth]{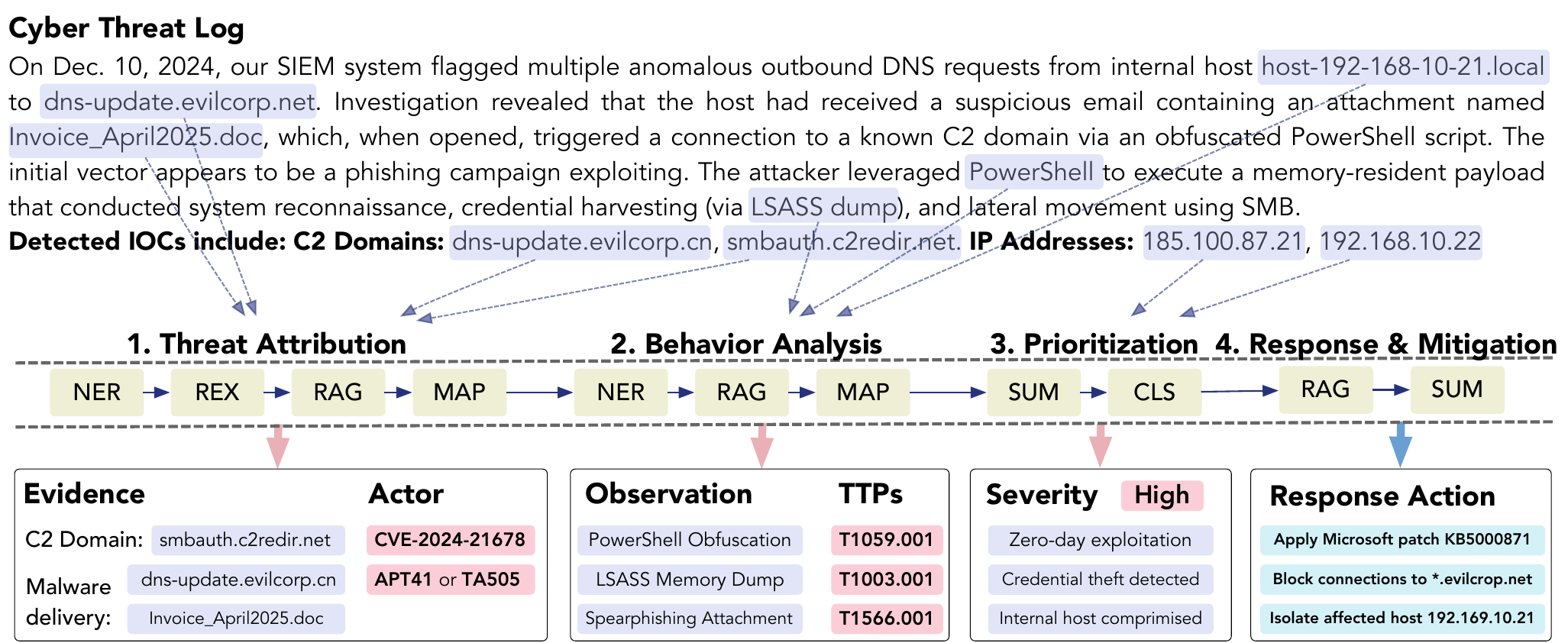}
    \caption{A threat hunting example equipped with the embodied functions. Function names: NER--named entity recognition, REX--regex parsing, MAP--text mapping, RAG--retrieval-augmented generation, CLS--classification, SUM--summarization.}
    \label{fig:illustration}
\end{figure}

{\bf Broader Coverage.} \system is constructed from a diverse and large-scale repository of threat intelligence data sourced from \ndb vulnerability databases, including MITRE series \cite{mitre_attack}, NVD \cite{nvd}, Exploit-DB \cite{exploitdb}, VulDB \cite{vuldb}, and CISE \cite{cise}, as well as reporting platforms such as Red Hat Bugzilla \cite{bugzilla}, Oracle Security Alerts \cite{oracle_security_alerts}, and IBM X-Force \cite{ibm_xforce}. This broad and representative collection is essential for capturing the complexity and variability of the modern threat landscape, enabling realistic support for both threat investigation and incident response. In addition, \system includes a significantly larger number of tasks and samples than existing cybersecurity benchmarks \cite{jimenez2023swe,li2025iris,alam2024ctibench,ji2024sevenllm}, as summarized in Table~\ref{tab:compare}. This extensive coverage allows for a more comprehensive and nuanced evaluation of LLM performance across a wide range of threat-hunting scenarios.

{\bf Embodied Environment.} An important feature of \system is its structured, modular workflow for guiding LLMs within an embodied environment \cite{yang2024holodeck, cheng2025embodiedeval}. This design is inspired by blue team practices, where analysts typically follow standardized procedures to interpret threat metadata and conduct investigations \cite{sehgal2023cybersecurity,diogenes2018cybersecurity,brotherston2024defensive}. However, strict adherence to such procedures can limit adaptability when analyzing unstructured threat logs or addressing emerging, zero-day threats.
To balance standardization and adaptability, \system integrates a set of embodied functions that regulate LLM behavior while allowing for open-ended reasoning where needed. As illustrated in Figure~\ref{fig:illustration}, \system first models the dependency structure among threat-hunting objectives (e.g., attribution, behavior analysis, mitigation) as a task chain, and then maps this chain into a corresponding sequence of embodied functions. In this process, functions such as NER enforce structured outputs (e.g., extracting threat actor entities), while functions like RAG support more flexible reasoning (e.g., summarizing relevant patching strategies).
This design enables \system to constrain model behavior when precision is essential, while still allowing LLMs to engage in context-sensitive generation when analytical creativity or abstraction is required.

{\bf Evaluation Strategy.}  \system incorporates agent-based evaluation strategies tailored to each threat-hunting objective. We benchmark leading LLMs and state-of-the-art (SOTA) cybersecurity agents, comparing \system’s embodied function-calling approach with popular elicitation strategies such as In-Context Learning (ICL) \cite{dong2022survey}, Chain-of-Thought (CoT) \cite{wei2022chain}, Tree-of-Thought (ToT) \cite{yao2023tree}. Our evaluation provides insights into the capabilities and limitations of current models across \ntask tasks. We further analyze LLMs' capabilities in automatically identifying task dependencies, selecting appropriate embodied functions, and handling noisy threat log inputs. These experiments reveal not only underlying model limitations, such as reasoning errors and sensitivity to incomplete or ambiguous input, but also provide practical insights for blue team analysts aiming to integrate LLMs into real-world cyber defense workflows.


In summary, this paper makes the following contributions:
(1) We introduce \system, a practice-informed, broadly scoped benchmark that enables rigorous evaluation of LLMs for blue team threat hunting, (2) we construct an embodied environment that models the dependencies among threat-hunting tasks and guides LLMs through standardized yet flexible reasoning workflow, (3) we conduct comprehensive evaluations and provide insights to improve LLM performance among threat-hunting scenarios.

\begin{table}[t]
\caption{Comparison of cybersecurity benchmarks for LLMs.}
\resizebox{\textwidth}{!}{
\centering
\small
\label{tab:compare}
\begin{tabular}{llcccll}
\toprule
\textbf{Benchmark} & \textbf{Focus} & \textbf{\#Data} & \textbf{\#Task} & \textbf{\#Source} & {\bf Coverage} & {\bf Unique Feature} \\
\lightmidrule
CWE-Bench-Java \cite{li2025iris} & Java vulnerability  & 120 & 4 & 1 & Four CWE classes & Large-scale Java codes\\
CTIBench \cite{alam2024ctibench} & Cyber Threat Intelligence & 2,500 & 3 & 6 &  CVE, CWE, CVSS, ATT\&CK & Multi-choice questions (MCQ)  \\
SevenLLM-Bench \cite{ji2024sevenllm} & Report understanding & 91,401 & 28 & N/A & Bilingual instruction corpus  & Synthetic Data, MCQ, QA\\
SWE-Bench \cite{jimenez2023swe} & Software bug fixing & 2,294 & 12 & 1 & GitHub issues & Python repository \\
\lightmidrule
\textbf{\system (Ours)} & Blue-team threat hunting & 452,293 & \ntask & \ndb & Threat-hunting lifecycle (\ref{ssec:task}) & Open Generation, Embodied Env  \\
\bottomrule
\end{tabular}
}
\end{table}

\section{Related Work}

{\bf LLMs for Cybersecurity.}
Recently, LLMs have shown promise in enhancing cybersecurity tasks such as malware classification \cite{abusitta2021malware, al2024exploring, qian2025lamd, devadiga2023gleam, liu2025cyber}, code vulnerability detection \cite{russell2018automated, lu2024grace, sheng2024lprotector}, penetration testing \cite{happe2023getting, muzsai2024hacksynth, shen2024pentestagent}, phishing detection \cite{kulkarni2024ml, greco2024david}, and incident report generation \cite{bernardi2024automatic, sufi2024innovative, mcgregor2025err}. These applications leverage the language understanding and reasoning capabilities of LLMs to analyze technical data, recommend solutions, or simulate attacker behaviors. However, existing applications typically target isolated tasks without considering broader analyst workflows. Additionally, their open-ended reasoning often results in hallucinations and inconsistencies \cite{mundler2023self, simhi2025trust, shrivastavaresponse}, raising concerns about reliability in high-stakes defensive scenarios.

{\bf Cybersecurity Benchmarks.} Recent benchmarks have focused on static analysis \cite{reinhold2024surmounting, higuera2020benchmarking, braga2017practical}, software vulnerabilities \cite{hossen2024assessing, sawant2024improving}, and threat report generation \cite{tihanyi2024cybermetric, perrina2023agir, vcupka2023comparison}. These benchmarks evaluate predefined tasks such as identifying CWE categories, matching CVEs, or summarizing intelligence reports \cite{alam2024ctibench, aghaei2020threatzoom, branescu2024automated, hemberg2020linking}. While helpful for reproducibility, they often cover narrow domains and lack the complexity and task interdependencies inherent in real-world threat investigations. In contrast, benchmarks from other high-stakes fields (e.g., law, medicine, finance) increasingly include complex, multistep tasks requiring diverse reasoning skills \cite{fei2023lawbench, wang2024mmlu, choshen2024navigating, lucas2024reasoning, zhourevisiting}. Inspired by these efforts, we introduce \system to emphasize structured reasoning and realistic interdependencies, specifically for blue team threat hunting.

{\bf Embodied Agents.} 
Recent research has proposed embodied agents and function-calling frameworks to structure LLM reasoning into modular, interpretable steps \cite{driess2023palm, dongre2024respact, hu2024agentgen}. Such frameworks have achieved notable success in robotics \cite{jeong2024survey, akkaladevi2021semantic}, database querying \cite{kadir2024systematic, dar2019frameworks}, and scientific reasoning tasks \cite{abate2020assessment, vaesen2021new}. However, their use in cybersecurity, especially defensive operations, remains underexplored despite the need for structured workflows. Our work addresses this gap by introducing an embodied, function-guided environment aligned with blue team practices, enabling procedural reasoning within a structured analytical pipeline.
\section{\system}
\label{sec:method}

In this section, we provide a detailed introduction of \system regarding the collected threat hunting tasks (\ref{ssec:task}), data sources (\ref{ssec:source}), and the embodied strategy (\ref{ssec:method}).

\subsection{Threat Hunting Tasks}
\label{ssec:task}

As shown in Table \ref{tab:benchmark},  \system reflects the full lifecycle of threat hunting tasks.  Specifically, \system systematizes analytical tasks into four categories: {\bf Threat Attribution}, {\bf Behavior Analysis}, {\bf Prioritization}, and {\bf Response \& Mitigation}. Each category captures a stage in the threat-hunting workflow from investigating cyber threats to identifying countermeasures. Specifically:

{\bf Threat Attribution} aims at uncovering the origins and nature of a threat. This includes tasks such as extracting infrastructure artifacts (e.g., domains, IPs, URLs), classifying malware families based on observed behaviors, matching known threat signatures, and linking activities to known campaigns or actor groups (e.g., APT or MITRE ATT\&CK \cite{mitre_attack}). Further granularity is achieved through geographic and temporal pattern analysis, as well as victimology and affiliation linking, all of which help analysts contextualize incidents in terms of their broader threat landscape. 

Subsequently, {\bf Behavior Analysis} focuses on understanding how adversaries interact with systems over time. Tasks in this category include mapping unusual file system activities, profiling network behaviors (e.g., Monitoring outbound traffic), detecting credential access, and analyzing the use of commands and scripts. Analysts aim to reconstruct sequences of attack events and associate them with specific execution contexts or behavioral patterns. The detection of techniques such as privilege escalation and defense evasion also falls within this scope. Understanding threat behaviors enhances an analyst's ability to assess the dynamics of emerging (or, zero-day) cyber threats.

\begin{table}[!t]
\caption{Threat hunting tasks, description of targets, corresponding embodied functions, number of instances, and evaluation metrics. We implememt \nfunc embodied functions: (1) NER: named entity recognition, (2) REX: regex parsing, (3) SUM: summarization, (4) SIM: text similarity matching, (5) MAP: text mapping, (6) RAG: retrieval-augmented generation, (7) SPA: text span localization, (8) CLS: classification, and (9) MATH: mathmatical calculation. We evaluate using metrics (i) F1 score, (ii) Sim: text similarity (by BERT Score \cite{bertscore}), (iii) Accuracy, (iv) Normalized distance between two numbers, (v) Passing rate of code execution, (vi) Hit@$k$ ratio. Details in Appendix \ref{app:embodied} and \ref{app:metric}.}
\resizebox{\textwidth}{!}{
\centering
\small
\begin{tabular}{lllll}
\toprule
\textbf{Task} & \textbf{Analytical Target} & \textbf{Function} & \textbf{\#Data}  & \textbf{Metric} \\
\lightmidrule
\rowcolor{yellowcolorone}
\multicolumn{5}{c}{\textbf{\textit{Threat Attribution}}}\\
\lightmidrule
Malware Identification & Malware delivery or toolset & NER, SUM & 15,742 & F1 \\ 
Signature Matching & Techniques from known threat groups & NER, SIM & 5,166 & F1 \\
Temporal Pattern Matching & Known work schedules & REX & 4,203 & Sim \\
Affiliation Linking & Source organizations & NER, MAP  & 17,583 & F1 \\
Geographic Analysis & Geographic or cultural indicators & NER, SIM & 6,164 & F1 \\
Victimology Profiling & Targeted victims or attacker motives & NER, REX  & 18,612 & F1 \\
Infrastructure Extraction & Domains, IPs, URLs, or file hashes & NER, REX, SUM & 24,129 & F1 \\
Actor Identification & The threat group or actor (e.g., APT28) & NER, RAG, MAP  & 17,823 & F1 \\
Campaign Correlation & Threat campaigns or incidents & NER, MAP & 27,762 & F1 \\ 
\lightmidrule
\rowcolor{yellowcolortwo}
\multicolumn{5}{c}{\textbf{\textit{Behavior Analysis}}}\\
\lightmidrule
File System Activity Detection & Suspicious file creation, deletion, or access & SPA, NER, SUM & 4,653 & Sim \\ 
Network Behavior Profiling & Patterns of external communication (e.g., C2) & SPA, NER, SUM & 2,617 & Sim \\ 
Credential Access Detection & Theft or misuse of credentials & SPA, NER, SUM & 2,492 & Sim \\ 
Execution Context Analysis & Execution behaviors by user or process & SPA, NER, SUM & 23,888 & Sim \\ 
Command \& Script Analysis & Suspicious commands or scripts & SPA, NER, SUM & 20,232 & F1 \\
Privilege Escalation Inference & Privilege escalation attempts & SPA, NER, SUM &  15,953 & Sim \\
Evasion Behavior Detection & Evasion or obfuscation techniques & SPA, NER, SUM & 8,973 & Sim \\
Event Sequence Reconstruction & Timeline of attack-related events & SUM & 23,265 & Sim \\
TTP Extraction & Tactics, techniques, and procedures & RAG, MAP & 28,292 & F1 \\
\lightmidrule
\rowcolor{purplecolor}
\multicolumn{5}{c}{\textbf{\textit{Prioritization}}}\\
\lightmidrule
Attack Vector Classification & Exploitation vectors (e.g., network, local, physical) & SUM, CLS & 17,448 & Acc \\
Attack Complexity Classification & Level of hurdles required to carry out the attack & SUM, CLS & 17,116 & Acc \\
Privileges Requirement Detection & Level of access privileges an attacker needs & SUM, CLS & 18,030 & Acc \\
User Interaction Categorization & If exploitation requires user participation  & SUM, CLS & 17,075 & Acc \\
Attack Scope Detection & If the vulnerability affects one/multiple components & SUM, CLS & 18,570 & Acc \\
Impact Level Classification & Consequences on confidentiality, integrity, and availability & SUM, CLS & 18,736 & Acc \\
Severity Scoring & A numerical score indicating the overall attack severity & SUM, MATH & 11,507 & Dist \\
\lightmidrule
\rowcolor{azure(colorwheel)!20}
\multicolumn{5}{c}{\textbf{\textit{Response \& Mitigation}}}\\
\lightmidrule
Playbook Recommendation &  Relevant response actions based on threat type & RAG, SUM & 10,718 & Hit \\
Security Control Adjustment & Firewall rules, EDR settings, or group policies & RAG, SUM & 9,929 & Sim \\
Patch Code Generation & Code snippets to patch the vulnerability & RAG, SUM & 11,341 & Pass \\
Patch Tool Suggestion & Security tools or utilities & RAG, SUM & 9,763 & Hit \\
Advisory Correlation & Security advisories or best practices & RAG, SUM & 24,511 & Hit \\
\bottomrule
\end{tabular}
}
\label{tab:benchmark}
\end{table}

When multiple threats emerge simultaneously, {\bf Prioritization} assesses their relative urgency and associated risk. This involves analyzing the attack vector and complexity, identifying privilege requirements and user interaction dependencies, and estimating potential impact. These factors are then synthesized into impact labels and severity scores (e.g., CVSS \cite{cvss}) to guide effective triage. Finally, {\bf Response \& Mitigation} focus on generating actionable defense strategies. This includes recommending response playbooks, generating patch code, correlating relevant security advisories, and suggesting appropriate tools or configuration changes to neutralize the threat.

\subsection{Data Sources}
\label{ssec:source}

\system collects threat metadata from two primary sources: (1) vulnerability databases, which offer authoritative structural and non-structural information about threats, and (2) threat intelligence platforms, which report event-driven, context-rich threat data.

{\bf Vulnerability databases} serve as foundational resources for automated threat hunting by providing machine-readable records of software flaws, exposure types, and critical contextual metadata. We aggregate threat entries from established sources such as NVD \cite{nvd}, MITRE CVE \cite{mitre_cve}, ATT\&CK \cite{mitre_attack}, CWE \cite{cwe}, CAPEC \cite{capec}, D3FEND \cite{d3fend}, Exploit-DB \cite{exploitdb}, and VulDB \cite{vuldb}. These sources include detailed insights such as exploitability scores (EPSS \cite{epss}), severity metrics (CVSS \cite{cvss}), and remediation guidance. Additionally, we incorporate data from vendor-maintained repositories (e.g., the Microsoft Security Update Guide \cite{microsoft_security}, IBM X-Force \cite{ibm_xforce}) to capture fine-grained details on affected systems, attack vectors, and patch methods.

\begin{figure}[!tp]
    \centering
    \includegraphics[width=\textwidth]{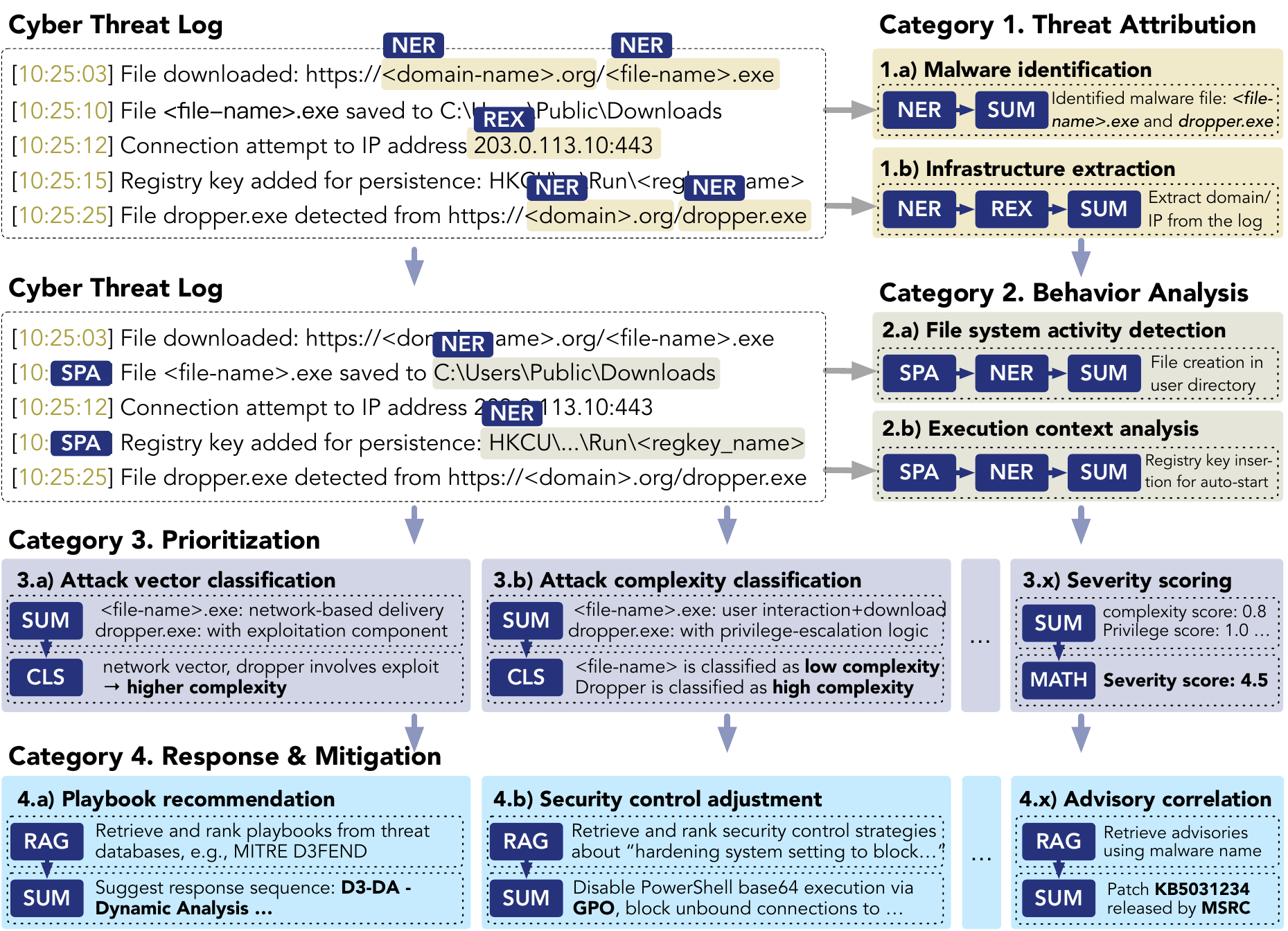}
    \caption{An threat-hunting example demonstrating a dependency chain of analytical tasks, where each task is completed through a sequence of embodied functions executed by LLMs autonomously.}
    \label{fig:dependency}
\end{figure}

{\bf Threat intelligence platforms} complement these databases by providing behavioral and contextual signals linked to adversary activity. Platforms such as VirusTotal \cite{virustotal}, AlienVault OTX \cite{alienvault_otx}, and MISP \cite{misp} contribute indicators of compromise (IOCs), behavioral patterns, and telemetry that enable tasks like campaign correlation, infrastructure extraction, and actor attribution. Furthermore, industry threat reports—from sources, such as Mandiant \cite{mandiant}, Recorded Future \cite{recordedfuture}, Palo Alto Unit 42 \cite{unit42}, and Apache \cite{apache}, offer semi-structured intelligence, including incident timelines, IOC lists, and narrative analyses, which are essential for modeling multi-stage attack sequences and evaluating blue team responses. 

Additional details on how these databases and platforms are used are provided in Appendix~\ref{app:source}.

\subsection{Embodied Threat Hunting: Definition and Methodology}
\label{ssec:method}

{\bf Task Dependency.} Threat hunting is inherently a multi-stage analytical process \cite{sauerwein2019, caltagirone2013diamond, hillier2022turning}, where downstream actions, such as incident response and mitigation, rely on outcomes derived from upstream analytical steps. For example, recommending an effective response playbook requires accurate attribution of the threat actor and thorough behavioral analysis of the compromise. To explicitly model this structured workflow, \system formulates threat hunting as a {\it Dependency Chain}. As illustrated in Figure \ref{fig:dependency}, all analytical tasks (e.g., 1.a: Malware Identification or 2.a: File System Activity Detection) are organized into a pipelined workflow that reflects their inherent dependencies. For example, attack complexity classification relies on prior analyses of file system activity and execution context. Meanwhile, tasks within the same category (e.g., malware identification and infrastructure extraction under threat attribution) can often be performed in parallel, as they address distinct dimensions of the threat and do not exhibit direct interdependencies.

{\bf Embodied Function.} Within each node, \system invokes a set of embodied functions designed to produce actionable threat intelligence and progressively address the current analytical target. Specifically, each threat hunting task \( t_i \) is associated with a corresponding set of tool-augmented functions \( \mathcal{F}_i = \{f_i^1, f_i^2, \dots \} \). We implement a total of \nfunc embodied functions, and solving each task \( t_i \) involves executing a predefined sequence \( f_i^* \in \mathcal{F}_i \), as detailed in the third column of Table~\ref{tab:benchmark}. The resulting output \( y_i = f_i^*(x) \) is subsequently passed to dependent downstream tasks. For instance, the task of {\it TTP Extraction} involves invoking both Retrieval-Augmented Generation (RAG) and Mapping (MAP) functions to identify relevant tactics and techniques from unstructured logs. Subsequently, a downstream task such as {\it Tool Suggestion} utilizes RAG and summarization (SUM) functions to map these identified TTPs to suitable defensive tools.

Due to space constraints, we defer implementation details embodied functions to Appendix \ref{app:embodied}.

\section{Experiment}
\label{sec:expt}

\system aims to empirically address the following research questions:
{\bf RQ$_1$:} How effective are embodied functions compared to standard prompting strategies (e.g., ICL, CoT, ToT) in improving LLM performance for threat-hunting tasks?
{\bf RQ$_2$:} Can LLMs accurately solve individual threat-hunting tasks?
{\bf RQ$_3$:} Are LLMs capable of reasoning about task dependencies and selecting the appropriate embodied functions for each task?
{\bf RQ$_4$:} How robust are LLMs, under the guidance of \system, when analyzing noisy inputs?

{\bf LLMs.} We evaluate a range of industry-leading large language models, including GPT-4o, GPT-o3, Qwen3-32B, Gemini-Pro, Claude-3-Opus, LLaMA-3.2-90B, LLaMA-3.1-405B, and DeepSeek-V3. In addition, we assess state-of-the-art cybersecurity-focused LLM agents, including Lily-Cybersecurity-7B \cite{LilyCybersecurity7B}, CyLens-8B \cite{liu2025cyber}, and SevenLLM-7B \cite{ji2024sevenllm}.

{\bf Elicitation (Prompting) Strategies.} To compare with the embodied function approach implemented in \system, we evaluate three widely used prompting strategies:
(1) In-Context Learning (ICL) \cite{dong2022survey} -- including basic task instructions along with five illustrative examples to demonstrate the desired solution format.
(2) Chain-of-Thought (CoT) \cite{wei2022chain} -- encouraging the model to generate "step-by-step" reasoning results before producing the final answer.
(3) Tree-of-Thought (ToT) \cite{yao2023tree} -- guiding LLMs to explore multiple reasoning paths and select the most plausible one. 


{\bf Metrics.} As shown in Table~\ref{tab:benchmark}, we select evaluation metrics tailored to the nature of each task. For information extraction tasks (e.g., malware identification), we use the {\bf F1 score} to balance precision and recall. For classification tasks (e.g., privilege escalation inference), we adopt {\bf accuracy} among well-defined categories. Generation or summarization tasks (e.g., behavioral profiling) are evaluated using {\bf BERTScore} \cite{bert-score}, reflecting semantic similarity. Tasks involving ranking (e.g., security playbook recommendation) utilize {\bf Hit@k} (default $k=10$), measuring whether correct choices appear in the top-k outputs. For programmatic outputs (e.g., patch code generation), we apply {\bf Pass} rate using \textsc{unitest} in Python to assess functional correctness. Numeric estimation tasks (e.g., severity scoring) are evaluated using {\bf Normalized Distance} to quantify similarity to ground truth values. All metrics are scaled to the range [0, 1] (or in percentages), where higher values indicate better performance. Additional details are provided in Appendix~\ref{app:metric}.

\subsection{Threat-Hunting Effectiveness through Embodied Functions (RQ$_1$)}
\label{ssec:expt-1}

\begin{table}[!t]
\caption{Results of LLMs threat-hunting performance (scaled to 100\%) on \system, using corresponding metrics tailored to each analytical target as detailed in Table \ref{tab:benchmark}. We use {\bf boldface} indicate the best results and \underline{underline} to denote the second-best results.}
\resizebox{\textwidth}{!}{
\centering
\begin{tabular}{l|cccc|cccc|cccc|cccc|cccc}
\toprule
\multicolumn{1}{c}{\multirow{2.5}{*}{\bf Model}} & 
\multicolumn{4}{c}{Playbook Recommend} & 
\multicolumn{4}{c}{Security Control Adjust} & 
\multicolumn{4}{c}{Patch Code Generation} & \multicolumn{4}{c}{Patch Tool Suggestion} & 
\multicolumn{4}{c}{Advisory Correlation}  \\
\cmidrule(lr){2-5}\cmidrule(lr){6-9}\cmidrule(lr){10-13}\cmidrule(lr){14-17}\cmidrule(lr){18-21}
 &  ICL & CoT & ToT & \hlcell Emb & ICL & CoT & ToT & \hlcell Emb & ICL & CoT & ToT & \hlcell Emb & ICL & CoT & ToT & \hlcell  Emb & ICL & CoT & ToT & \hlcell Emb \\
\midrule
\rowcolor{yellowcolorone!30}
\multicolumn{21}{c}{\it Cybersecurity Agent} \\
\midrule
Lily-7B & 42.3 & \underline{51.6} & 48.1 & \hlcell {\bf 67.2} & 51.5 & 60.3 & \underline{66.7} & \hlcell {\bf 74.2} & 10.8 & 24.5 & \underline{25.3} & \hlcell {\bf 29.7} & 48.2 & 53.6 & \underline{56.5} & \hlcell {\bf 69.1} & 21.7 & \underline{49.5} & 46.8 & \hlcell {\bf 73.4} \\
CyLens-8B & 83.5 & 80.6 & \underline{83.8} & \hlcell {\bf 88.2} & 75.2 & 80.6 & \underline{83.1} & \hlcell {\bf 87.5} & 62.4 & \underline{68.7} & 61.9 & \hlcell {\bf 80.8} & 74.6 & 80.3 & \underline{83.5} & \hlcell {\bf 88.9} & 66.5 & 82.1 & \underline{85.3} & \hlcell {\bf 89.8} \\
SevenLLM-7B & \underline{54.7} & 50.5 & 54.3 & \hlcell {\bf 66.8} & 43.9 & \underline{68.4} & 61.6 & \hlcell {\bf 80.1} & 29.2 & 55.1 & \underline{58.3} & \hlcell {\bf 60.2} & 61.5 & \underline{77.2} & 68.1 & \hlcell {\bf 77.7} & 63.8 & \underline{69.5} & 67.2 & \hlcell {\bf 77.1} \\
\midrule
\rowcolor{yellowcolorone!30}
\multicolumn{21}{c}{\it Industry-Leading LLM} \\
\midrule
GPT-4o & 64.5 & \underline{78.3} & 75.2 & \hlcell {\bf 84.6} & 61.8 & 70.3 & \underline{75.9} & \hlcell {\bf 82.1} & 56.2 & 58.4 & \underline{61.8} & \hlcell {\bf 72.5} & 68.9 & \underline{79.2} & 75.8 & \hlcell {\bf 87.4} & 64.7 & 67.2 & \underline{70.8}& \hlcell {\bf 80.3} \\
GPT-o3 & 73.1 & \underline{88.2} & 84.6 & \hlcell {\bf 90.2} & 70.3 & 79.5 & \underline{84.7} & \hlcell {\bf 88.3} & 58.4 & \underline{75.6} & 71.3 & \hlcell {\bf 86.9} & 79.4 & \underline{89.5} & 85.2 & \hlcell {\bf 96.3} & 67.2 & 79.8 & \underline{83.6} &\hlcell {\bf 91.7} \\
QWen3-32B & 52.8 & 67.5 & \underline{71.4} & \hlcell {\bf 79.3} & 50.6 & 59.8 & \underline{66.3} & \hlcell {\bf 74.7} & 39.3 & \underline{54.7} & 50.2 & \hlcell {\bf 65.4} & 59.2 & 70.3 & \underline{74.5} & \hlcell {\bf 83.6} & 48.5 & 61.7 & \underline{64.8} & \hlcell {\bf 76.5} \\
Gemini-Pro & 79.4 & 80.1 & \underline{83.5} & \hlcell {\bf 91.8} & 65.8 & \underline{79.2} & 73.6 & \hlcell {\bf 88.5} & 63.7 & 65.3 & \underline{69.8} & \hlcell {\bf 82.6} & 74.1 & 81.7 & \underline{86.3} & \hlcell {\bf 93.2} & 62.4 & \underline{77.5} & 73.1 & \hlcell {\bf 86.9} \\
Claude-Opus & 63.7 & \underline{80.6} & 76.4 & \hlcell {\bf 88.5} & \underline{79.2} & 76.3 & 72.1 & \hlcell {\bf 85.8} & 47.5 & \underline{65.2} & 60.8 & \hlcell {\bf 78.4} & 68.5 & 78.3  & \underline{82.9} & \hlcell {\bf 90.6} & 56.8 & \underline{75.1} & 71.3 & \hlcell {\bf 83.7} \\
Llama-90B & \underline{77.2} & 74.5 & 69.3 & \hlcell {\bf 82.6} & 53.4 & \underline{70.8} & 64.5 & \hlcell {\bf 79.3} & 42.8 & 56.2 & \underline{60.3} & \hlcell {\bf 71.5} & 64.1 & \underline{76.8} & 72.5 & \hlcell {\bf 85.2} & 51.3  & 63.7 & \underline{68.4} & \hlcell {\bf 77.8} \\
Llama-405B & 65.8 & 77.3 & \underline{82.1} & \hlcell {\bf 89.7} & 61.5 & \underline{77.9} & 72.8 & \hlcell {\bf 86.4} & 49.2 & \underline{67.4} & 62.9 & \hlcell {\bf 80.6} & 70.3 & 79.6 & \underline{84.2}& \hlcell {\bf 92.1} & 58.7 & \underline{76.3} & 71.8 & \hlcell {\bf 84.9} \\
DeepSeek-V3 & 61.4 & \underline{79.8} & 75.1 & \hlcell {\bf 87.2} & 57.6 & \underline{74.3} & 68.9 & \hlcell {\bf 83.5} & 45.7  & 59.8 & \underline{63.5} & \hlcell {\bf 76.2} & 67.1  & 76.9 & \underline{81.4} & \hlcell {\bf 89.3} & 54.2 & \underline{72.8} & 68.4 & \hlcell {\bf 82.6} \\
\bottomrule
\end{tabular}}
\label{tab:main-expt}
\end{table}

Ultimately, \system is designed to generate actionable responses and mitigation strategies against cyber threats. We begin by evaluating the overall quality of LLM-generated responses and mitigation outputs on \system. Table \ref{tab:main-expt} presents the results, using task-specific metrics detailed in Table \ref{tab:benchmark}. From these results, we observe that using embodied functions outperforms standard elicitation methods. For instance, embodied functions guide GPT-o3 to achieve over 90\% Hit@10 in playbook recommendation and over 91\% in advisory correlation. This demonstrates the effectiveness of combining modular, task-specific guidance with the inherent flexibility of LLMs.

Notably, while ICL, CoT, and ToT have been shown to improve generation quality for general-purpose tasks \cite{dong2022survey, yu2023towards, wang2022towards}, they lack meaningful guidance for domain-specific problems that require precise procedural knowledge and structured analytical workflows. 

\begin{redtheobox}
{\bf Case Study I} (Failure Case). When using CoT to generate a response plan for LockBit (a ransomware), GPT-4o offers generic recommendations {\it "... the first step is to isolate affected machines. Next, the system should assess backup availability and notify stakeholders ..."} without tailoring to LockBit and ignoring unique traits like double extortion tactics or known exploits. 
\end{redtheobox}

By contrast, embodied functions in \system constrain LLM reasoning to follow predefined task sequences, ensuring outputs remain aligned with operational goals:

\begin{theobox}
{\bf Case Study II} (Successful Case). The embodied function framework guides GPT-4o to explicitly invoke {\bf RAG} and {\bf SUM} modules. Specifically, RAG retrieves up-to-date security advisories (e.g., {\it CISA Alert AA23-325A}) specific to LockBit, while SUM outlines mitigation strategies with {\it double extortion prevention} and {\it air-gapped offline backups}.
\end{theobox}

These results suggest that in cybersecurity, particularly in threat-hunting scenarios, structured elicitation methods are necessary for reliably leveraging LLM capabilities.

\subsection{Threat-Hunting Performance for Individual Tasks (RQ$_2$)}

\begin{figure}[!tp]
    \centering
    \includegraphics[width=\textwidth]{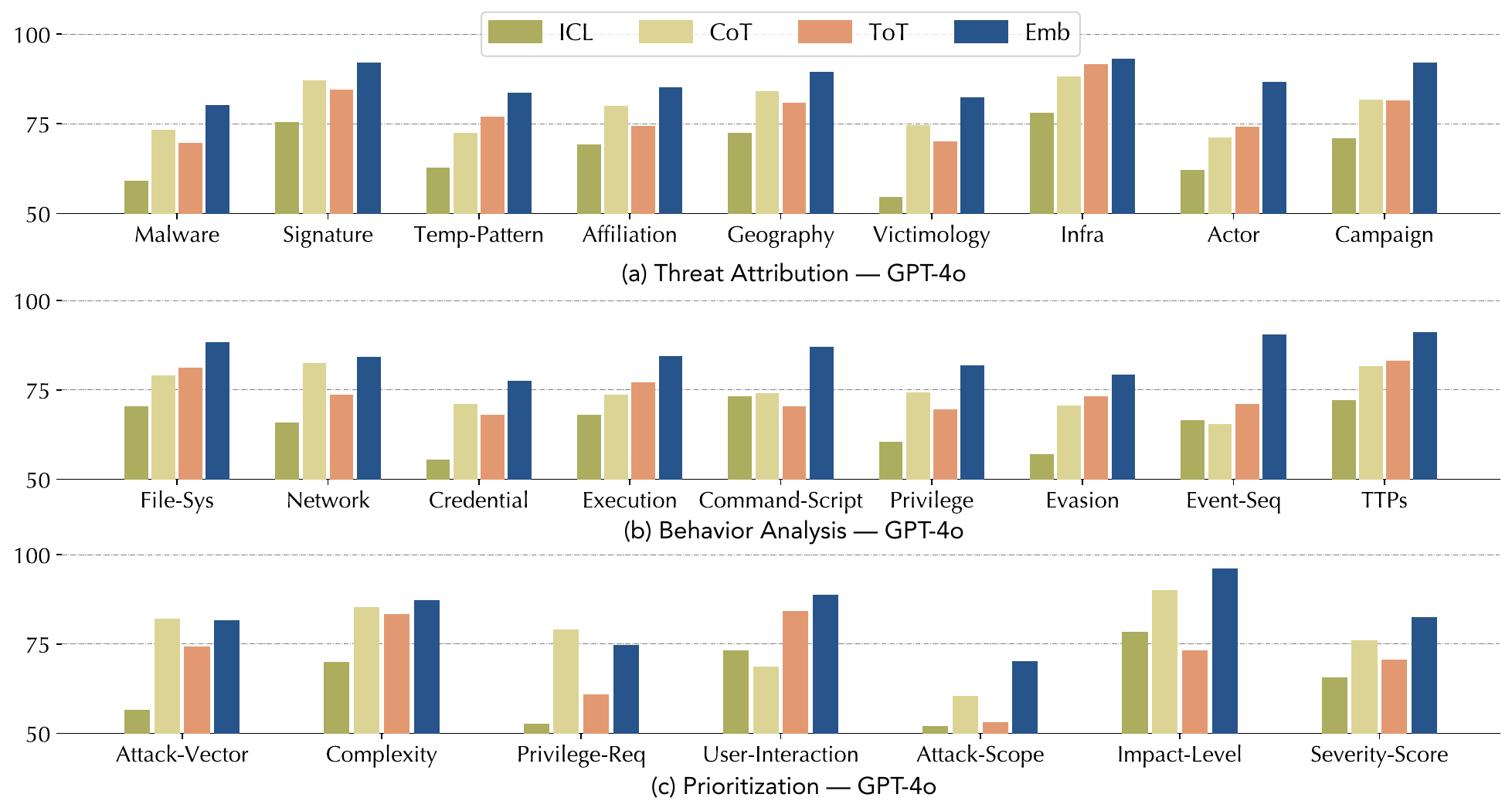}
    \caption{Threat-hunting performance (scaled to 100\%) on individual tasks, evaluating under GPT-4o across various elicitation strategies: ICL, CoT, ToT, and using embodied functions (Emb). Results for additional LLMs are provided in Appendix \ref{app:expt}.}
    \label{fig:expt-task}
\end{figure}

Complementing Section~\ref{ssec:expt-1}, we also evaluate individual threat-hunting tasks prior to the response \& mitigation stage, as outlined in Table~\ref{tab:benchmark}. Figures~\ref{fig:expt-task} and Appendix \ref{app:expt} present the experimental results.

Observe that using embodied functions consistently achieves the highest performance across all intermediate tasks. However, {\bf the magnitude of performance gains varies across task types}. For instance, in complex reasoning tasks (e.g., Event Sequence Construction), embodied functions yield substantial improvements over baseline strategies like CoT and ToT, boosting accuracy by over 20\% using GPT-4o. This is largely because these tasks require multi-hop reasoning, evidence synthesis, and careful dependency tracking, which are capabilities that general prompting methods struggle to coordinate effectively. In contrast, for narrower, classification-focused tasks (e.g., attack vector categorization or privilege escalation inference), the performance gap between embodied functions and standard prompting is smaller. Here, the tasks are more self-contained, and models can often arrive at correct predictions even without explicit task decomposition or function integration.

These findings highlight that while embodied functions offer general advantages, their relative benefit is particularly significant in scenarios requiring structured reasoning over interconnected threat-hunting steps. This demonstrates the importance of modular guidance in complex threat-hunting workflows.

\subsection{Task Dependency Resolution and Embodied Function Selection (RQ$_3$)}
\label{ssec:expt-3}

Previous experiments highlight the importance of guided workflows in threat hunting. Here, we further investigate LLMs’ inherent abilities to (1) identify which prior tasks provide the necessary inputs for solving a given analytical target, and (2) determine which embodied function(s) to invoke for executing specific threat-hunting tasks.

\begin{figure}[t]
    \centering
    \includegraphics[width=\textwidth]{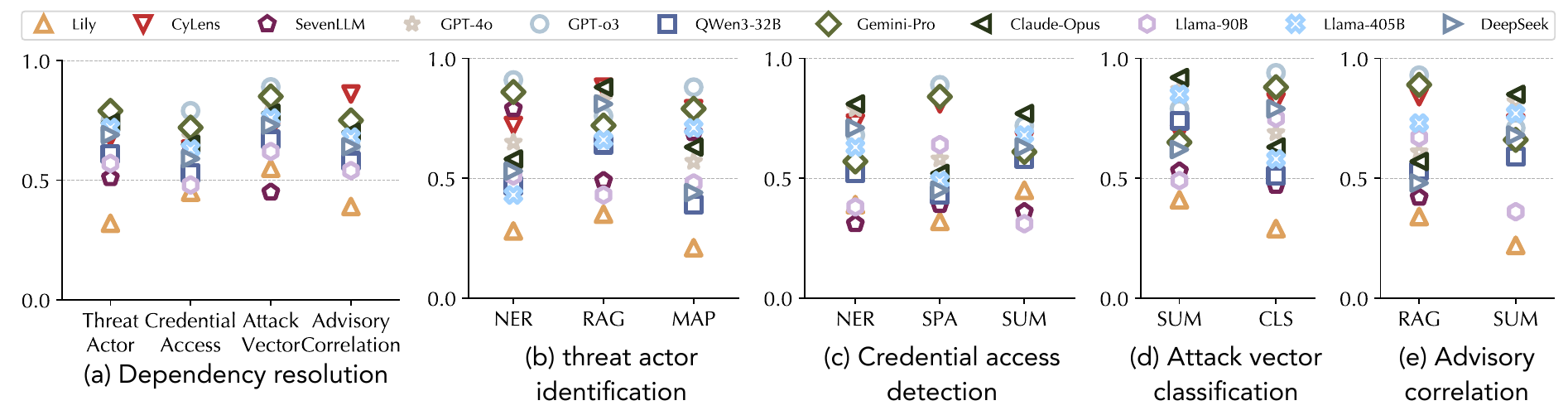}
    \caption{Evaluation of LLM performance in selecting the correct (a) task dependencies, and (b–e) embodied functions for specific analytical targets.}
    \label{fig:expt-select}
\end{figure}

{\bf Experimental Setting.} We frame both evaluations as {\it multi-choice problems}. For task dependency resolution, we present original threat logs from \system and prompt LLMs to identify which prior tasks must be resolved before proceeding with the current one. For example, when performing threat actor identification, LLMs are asked to select dependencies that provide necessary supporting evidence (e.g., signature extraction or affiliation mapping). The candidate options are drawn from the full set of tasks listed in Table~\ref{tab:benchmark}. For embodied function selection, LLMs are asked to choose the correct function(s) required at each step of a function-calling workflow. The candidate options are drawn from the complete set of \nfunc embodied functions defined in our system.

We measure the {\bf F1 score} for correct dependent task selection and {\bf accuracy} for correct function selection at each step. Our evaluation focuses on four representative tasks spanning the end-to-end workflow, from threat attribution to response \& mitigation.

{\bf Results and Observations.} Figure~\ref{fig:expt-select} presents the evaluation results. Observe that LLMs are largely insufficient in resolving task dependencies and accurately selecting the correct embodied functions without structured guidance. Specifically, while GPT-o3 performs better in general, it only achieves 64\% on identifying prior tasks required for advisory correlation. Open-source models such as Lily-7B exhibits even lower performance, often failing to recognize multi-step dependencies or indirect task linkages, leading to performance below 50\%.

For embodied function selection, the results show a similar trend: Gemini-Pro and CyLens achieve average accuracy over 75\%, while Llama-90B struggles to exceed 50\%, particularly on complex tasks such as threat actor identification or advisory correlation. Error analysis reveals two main failure cases: (i) {\it over-selection}, where the model includes irrelevant functions, and (ii) {\it under-selection}, where the model omits necessary functions required to complete the analytical workflow.

\begin{redtheobox}
{\bf Case Study III}  (Failure Case). Hallucinative Embodied Function Selection.

\vspace{3pt}
{\bf (i) Over-selection:} In {\it advisory correlation}, we observe that {\it Llama-90B} over-selects the irrelevant  {\it MAP} function, redundantly performing entity-to-infrastructure mapping. 

\vspace{3pt}
{\bf (ii) Under-selection:} In the {\it threat actor attribution}, {\it Claude} omits {\it RAG} (necessary to retrieve actor identifiers)
but directly generating APT identifiers, leading to an incomplete attribution.
\end{redtheobox}
    
These observations highlight the necessity of explicitly modeling multi-step dependency resolution and modular workflows. \system integrates these components to substantially mitigate the identified weaknesses, enabling LLMs to navigate threat-hunting pipelines with greater reliability.

\begin{figure}[t]
    \centering
    \includegraphics[width=\textwidth]{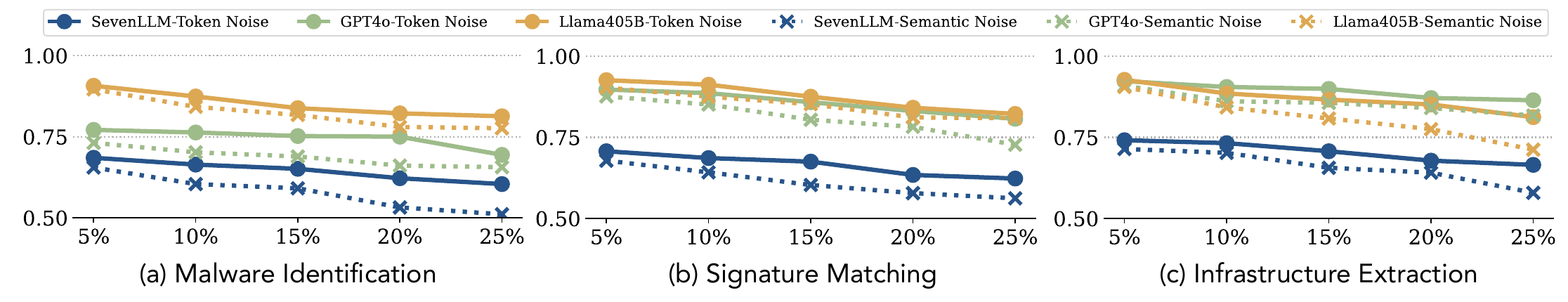}
    \caption{LLM performance when input threat logs are perturbed with token-level noise (solid line) or semantic-level noise (dashed line).}
    \label{fig:expt-noise}
\end{figure}

\subsection{LLM Robustness against Noisy Inputs (RQ$_4$)}

{\bf Experimental Setting.} We also investigate LLM robustness when input threat logs contain noisy text. We introduce
(i) token-level noise using TextAttack \cite{morris2020textattack}, which randomly injects or substitutes tokens, and
(ii) semantic-level noise using BART-paraphraser \cite{lewis2019bart}, which subtly introduces misleading or shifted context.
Both noise types are applied at controlled levels (e.g., perturbing 10\% of the input).


{\bf Results and Observations.} From Figure~\ref{fig:expt-noise}, we observe that token-level noise has a smaller impact on LLM performance compared to semantic-level noise. For example, under 10\% perturbation, random character insertions or deletions lead to less than 5\% performance drop across tasks. In contrast, semantic-level noise—such as paraphrased or subtly altered context—causes a much larger decline. These findings suggest that while LLMs handle surface-level errors relatively well, they struggle when the semantic shifting, even when guided by \system. This highlights the importance of curating expert-level threat reports in threat hunting, as imprecise statements can unintentionally mislead blue team efforts and degrade overall analysis.


\section{Conclusion}

We present \system, a benchmark designed to evaluate the capabilities of LLMs in blue team threat-hunting workflows. By combining broad and diverse real-world datasets, an embodied environment with modular function-guided reasoning, and detailed evaluation strategies, \system provides a comprehensive workflow for assessing LLM capabilities in realistic cyber defense scenarios. Our empirical findings highlight both the strengths and current limitations of SOTA LLMs, offering actionable insights for improving their integration into security operations. We hope \system will serve as a valuable resource for the research community and practitioners alike, driving future innovations in AI-assisted cybersecurity.

\newpage 
\bibliographystyle{plain}
\bibliography{neurips_2025.bib}
\label{reference}

\newpage
\appendix
\section{Data Source and Metadata Collection}
\label{app:source}

\textbf{The MITRE CVE (Common Vulnerabilities and Exposures)} system \cite{mitre_cve} is a foundational database that provides unique identifiers for publicly disclosed cybersecurity vulnerabilities. Each CVE record includes an ID, a brief description, references to external resources, and associated vendors or platforms. This source allows for consistent naming and indexing of vulnerabilities across tools and reports.
We collect structured metadata such as CVE IDs, descriptions, reference links, and related CWE classifications. CVE feeds (XML/JSON) are used for automated ingestion and linkage to other threat intelligence frameworks like CAPEC and ATT\&CK.

Maintained by NIST, the \textbf{NVD (National Vulnerability Database)}  \cite{nvd} builds on MITRE CVE data by adding rich metadata, including CVSS scores (base, temporal, environmental), CWE mappings, configuration impacts, patch availability, and severity vectors.
We extract metadata through the official JSON data feeds, parsing CVE-level risk metrics, impact sub-scores, and associated product configurations. This information is critical for prioritizing remediation and understanding the real-world impact of vulnerabilities.

\textbf{Exploit-DB}  \cite{exploitdb} is a curated collection of publicly available exploits and proof-of-concept code. Each entry includes exploit titles, CVE references, author information, platform tags, and the actual code used in attacks.
Unlike CVE/NVD, Exploit-DB provides practical insights into how vulnerabilities are weaponized in real environments. We extract titles, descriptions, exploit types (e.g., Local, Remote), and related CVEs using web scraping and NLP-based text classification.

\textbf{CWE(Common Weakness Enumeration)} \cite{cwe} is a taxonomy developed by MITRE to classify software and hardware weaknesses. Each CWE includes a unique ID, a detailed explanation, potential consequences, examples, and related patterns (e.g., CAPEC).      We use CWE to enrich CVE data with root cause information, enabling fine-grained vulnerability clustering and defensive prioritization. Metadata includes weakness category, severity, and relationships with CAPEC and CVE entries.

\textbf{CAPEC (Common Attack Pattern Enumeration and Classification)} \cite{capec} provides a standardized catalog of common attack strategies. Each pattern includes the attacker’s objectives, prerequisites, execution flow, related weaknesses (CWE), and example scenarios.  
We extract attack pattern IDs, descriptions, related CWEs, and suggested mitigations. These data points enable us to map vulnerabilities to adversarial behaviors, enhancing our CTI behavioral modeling capabilities.

\textbf{The MITRE ATT\&CK}  \cite{mitre_attack} framework systematically catalogs adversary tactics, techniques, and procedures (TTPs) observed in real-world incidents. Each entry includes tactic categories (e.g., Privilege Escalation), techniques, mitigations, detection suggestions, and threat actor mappings.  
We extract technique IDs, corresponding software, mitigation strategies, and detection methods. These are used to link CVEs and exploits to higher-level attacker behaviors, supporting advanced threat modeling.

{\bf D3FEND} \cite{d3fend} is a curated knowledge graph that maps defensive techniques to specific threat behaviors and artifacts. D3FEND complements the well-known ATT\&CK framework by focusing on how defenders can detect, disrupt, and respond to adversarial actions. To integrate this resource into \system, we crawl D3FEND’s publicly available ontology and extract metadata on detection, deception, and mitigation techniques, along with their associated digital artifacts (e.g., file paths, registry keys, network signatures). This metadata is then linked to relevant analytical tasks—such as behavioral profiling and response planning—providing a rich, standardized reference for grounding LLM outputs in practical defensive actions.

\textbf{Oracle Security Alerts}  \cite{oracle_security_alerts} provides detailed security patch advisories for its product suite. Each alert includes the CVEs addressed, severity scores, and remediation timelines.  
We parse the advisories to gather product-specific vulnerability timelines, vendor patch statuses, and mitigation instructions, which complement the NVD and MITRE CVE datasets.

\textbf{Red Hat Bugzilla}  \cite{bugzilla} is a bug tracking system that includes detailed discussions and technical logs about software bugs, many of which are security-related. Entries often include CVE links, fix status, patch availability, and affected components. 
We scrape metadata such as Bug IDs, CVE references, affected packages, and resolution details to supplement our understanding of vulnerability lifecycle management.

\textbf{The RHSA(Red Hat Security Advisories)} \cite{redhat_cve} portal lists all critical, important, and moderate security advisories affecting Red Hat products. Each advisory provides CVE mappings, severity scores, fixed packages, and risk summaries.  
Metadata extraction includes advisory IDs, publication dates, CVE linkages, and suggested upgrades or patches, enabling alignment with real-world remediation practices.

\textbf{IBM X-Force Exchange} \cite{ibm_xforce} is a commercial threat intelligence sharing platform that provides in-depth reports on vulnerabilities, exploits, malware, and threat actors. Each CVE entry is enriched with exploitability status, malware connections, and actor attribution.  
We extract structured threat metadata such as exploit availability, indicators of compromise (IOCs), campaign tags, and actor profiling to complement CVE risk modeling.

\textbf{CISE (Cybersecurity Information Sharing Environment)}  \cite{cise}, maintained by CISA, promotes cybersecurity information exchange across government and private sector entities. The platform facilitates sharing of indicators of compromise (IOCs), analysis reports, and threat mitigation strategies through structured partnerships.  
We extract strategic-level threat metadata, including threat vectors, vulnerability trends, and response best practices from shared reports and alerts. This supports broader CTI tasks like attribution and risk contextualization.

\textbf{VulDB (Vulnerability Database)} \cite{vuldb} is a commercial vulnerability intelligence service that provides insights into current exploits, threat actor behavior, and exploit trends. Entries often include exploitability scores, attack vectors, exploitation status, and tags related to malware or campaigns.  
We collect CVE mappings, vulnerability titles, exploitation timelines, and associated actors, enabling temporal and behavioral correlation with other sources like Exploit-DB and MITRE ATT\&CK .

\textbf{Apache}’s official security advisory page lists all disclosed vulnerabilities affecting Apache projects (e.g., HTTP Server, Tomcat, Struts) \cite{apache}. Each advisory includes CVE references, affected versions, and patch instructions.  
We extract CVE mappings, patch details, vulnerability types, and affected modules. These insights are cross-referenced with MITRE CVE and NVD entries to improve accuracy in software-specific threat tracking.

\textbf{Mandiant Threat Intelligence Reports} \cite{mandiant}, now part of Google Cloud, publishes in-depth research on nation-state APTs, malware campaigns, and threat actor tactics. Their reports include IOC lists, ATT\&CK mappings, and campaign chronologies.  
We extract metadata on APT groups, attack stages, observed TTPs, and malware toolkits. These data points support the attribution and behavioral modeling dimensions of our threat intelligence corpus.
    
\textbf{Recorded Future Threat Intelligence Reports} \cite{recordedfuture} publishes real-time, machine-readable threat intelligence covering threat actors, vulnerabilities, dark web chatter, and geopolitical cyber campaigns. Reports often include structured indicators, predictive analytics, and CVE exploitability assessments.  
We leverage this source to collect threat context, emerging trends, and exploit discussion patterns—enabling our system to associate vulnerabilities with evolving threat actor intent and capability .

\textbf{Unit 42 Threat Research (Palo Alto Networks)} \cite{unit42} provides malware analysis, campaign forensics, and actor behavior insights from Palo Alto Networks' global threat intelligence platform. Their publications include links to malicious infrastructure, malware families, and ATT\&CK references.  
We extract TTPs, CVE-to-malware correlations, and campaign data. This enhances our contextual metadata for linking specific vulnerabilities to real-world exploitation scenarios .

\textbf{Microsoft’s Security Update Guide} \cite{microsoft_security} lists monthly updates across its software stack. Entries contain CVEs, severity ratings, exploitability assessments, patch availability, and affected platforms.  
Metadata extraction includes CVE linkage, threat vectors (e.g., local, remote), exploitation likelihood, and patch rollout status—enriching vendor-specific vulnerability intelligence .

\textbf{CVSS (Common Vulnerability Scoring System)}  \cite{cvss} is a widely adopted scoring system developed by FIRST to assess the severity of software vulnerabilities. It breaks down risk into Base, Temporal, and Environmental components.  
We use this framework to interpret NVD scores, compare severity across platforms, and calibrate exploitability in relation to business-critical systems.

\textbf{EPSS (Exploit Prediction Scoring System)} \cite{epss}, also developed by FIRST, provides probabilistic predictions of whether a vulnerability is likely to be exploited in the wild. It integrates data from CVSS, Exploit-DB, and historical attack patterns.  
We ingest EPSS scores via API to prioritize vulnerabilities not just by severity, but by real-world exploitation likelihood—enabling dynamic risk-based vulnerability management.

\textbf{MISP (Malware Information Sharing Platform)}  \cite{misp} is an open-source platform designed for structured threat intelligence sharing using STIX/TAXII formats. It facilitates sharing of IOCs, threat event correlations, and TTP mappings.  
We integrate MISP data via its API to ingest indicators (e.g., hashes, domains, IPs), related threat actors, and event metadata. These enrich our knowledge graph with actionable CTI feeds.

{\bf VirusTotal} \cite{virustotal} is a widely used threat intelligence platform that aggregates malware analysis and sandbox reports from multiple antivirus engines and security vendors. To support behavior analysis and attribution tasks, \system collects structured threat metadata from VirusTotal’s public API, including file hashes (MD5, SHA-1, SHA-256), behavioral execution traces, contacted IPs/domains, dropped files, and detection labels. This information is linked to threat artifacts such as malware families, indicators of compromise (IOCs), and known campaign signatures. The extracted metadata enables \system to contextualize adversarial behaviors and enrich analytical functions like malware classification, infrastructure extraction, and campaign correlation.

{\bf AlienVault Open Threat Exchange (OTX)} \cite{alienvault_otx} is a collaborative threat-sharing platform that provides community-contributed threat indicators and contextual threat intelligence. \system leverages the OTX API to collect threat pulses—curated collections of IOCs and metadata describing specific threat actors, campaigns, or vulnerabilities. These pulses include information such as associated IPs, domains, file hashes, CVEs, and targeted sectors. By integrating OTX data, \system enhances its ability to support tasks like actor attribution, TTP matching, and community correlation, allowing LLMs to reason over shared intelligence and align analysis with ongoing threat landscapes.

\section{Embodied Functions: Complementary Detail}
\label{app:embodied}

To support modular and extensible capabilities within our \system, we decompose complex NLP workflows into discrete, embodied functions. This section detail the implementation of twelve NLP modules as described in section \ref{ssec:task}. Each function corresponds to a specific operation type, described as follows:

\subsection{NER (Named Entity Recognition)}

To identify and classify cybersecurity-relevant entities such as threat actors, malware names, vulnerabilities, and indicators of compromise (IOCs) in unstructured textual data, NER facilitates automated extraction for threat attribution and situational awareness. We employ prompt-based techniques that enable entity recognition without retraining, thus maintaining adaptability to emerging domain vocabulary.

\begin{tcolorbox}[title=Prompt 1. NER Prompt for Threat Attribution]
\textbf{System Prompt:}  
You are a cybersecurity threat intelligence assistant specialized in named entity recognition. Your task is to extract and categorize all named entities relevant to threat attribution from the provided text. Focus on answering:  
\textit{"Who is responsible for the attack?"},  
\textit{"How was the attack carried out?"}.

\textbf{Instructions:}  
Given a cybersecurity-related document or report excerpt, extract all relevant named entities and classify them into:

\begin{itemize}
  \item \textbf{Threat Actor:} Individual(s) or groups suspected or known to conduct the activity.
  \item \textbf{Malware/Tool:} Names of malicious software, exploits, or hacking tools.
  \item \textbf{Vulnerability:} CVE identifiers or technical flaws exploited.
  \item \textbf{Infrastructure:} IPs, domains, file hashes, or URLs used.
\end{itemize}

\textbf{Output:}  
Return results as a structured JSON object.
\end{tcolorbox}

\subsection{REX (Regex Parsing)}

To extract structured indicators from cybersecurity logs or reports, REX employs predefined regular expressions to match patterns like IP addresses, domain names, file hashes, and timestamps. This rule-based approach offers high precision in normalizing threat data for correlation and enrichment tasks.

\begin{tcolorbox}[title=Prompt 2. Regex Pattern Matching Prompt]
\textbf{System Prompt:}  
You are a cybersecurity parsing assistant. Your task is to extract standard threat indicators from raw incident reports using predefined regex patterns.

\textbf{Instructions:}  
Parse the following document and extract any matches for:
\begin{itemize}
  \item IP addresses
  \item File hashes (MD5, SHA1, SHA256)
  \item Domain names
  \item Timestamps
\end{itemize}

\textbf{Output:}  
Return all matches grouped by type in structured JSON format.
\end{tcolorbox}

\subsection{SUM (Summarization)}

To enable analysts to quickly grasp key information from lengthy threat reports, SUM generates concise summaries while preserving critical details such as TTPs, IOCs, and incident timelines.

\begin{tcolorbox}[title=Prompt 3. Threat Report Summarization Prompt]
\textbf{System Prompt:}  
You are a cybersecurity analyst assistant. Your task is to summarize the following threat report in 3–4 sentences, preserving the attack vector, affected systems, timeline, and any mentioned threat actors or IOCs.

\textbf{Instructions:}  
Summarize only the essential intelligence. Avoid generic phrases. Include dates, names, and tools where available.

\textbf{Output:}  
Return a plain-text summary paragraph.
\end{tcolorbox}

\subsection{SIM (Text Similarity Matching)}

To determine semantic equivalence between pairs of threat indicators—particularly geographic or cultural references (e.g., "Eastern European" vs. "Russian-speaking")—the SIM function applies LLM-based textual similarity matching. This is critical for normalizing contextual descriptions found in incident reports or threat assessments that use varied, informal, or aliasing terms to describe similar threat origin profiles. Rather than relying on surface-level keyword overlap, SIM leverages the LLM’s contextual understanding to judge whether two descriptions refer to the same underlying group or region. This helps unify disparate threat intelligence entries that may use different terminology for the same adversarial origin.

\begin{tcolorbox}[title=Prompt 4. Text Similarity Matching Prompt for Geocultural Indicators]
\textbf{System Prompt:}
You are a cybersecurity assistant that helps analysts determine whether two geolocation or cultural indicators refer to the same threat origin. Use contextual reasoning to decide if the two phrases describe the same group or region in a cyber threat context.

\textbf{Instructions:}
Given two input phrases describing threat origin (e.g., "Russian-affiliated" vs. "Eastern Bloc actor"), determine whether they semantically refer to the same group or geopolitical background.

Answer the following questions:
\begin{itemize}
\item Do both descriptions point to the same cultural, linguistic, or geopolitical region?
\item Are the expressions used interchangeably in threat intelligence contexts?
\end{itemize}

\textbf{Output:}
Return a JSON object with:
\begin{itemize}
\item \texttt{"match"}: Boolean (true/false)
\item \texttt{"confidence"}: A float score from 0.0 to 1.0
\item \texttt{"justification"}: One or two sentences explaining the decision
\end{itemize}
\end{tcolorbox}

\subsection{MAP (Text Mapping)}

To visualize and semantically relate named entities and key concepts extracted from cybersecurity documents, the \textbf{MAP} function supports construction of structured representations such as knowledge graphs or threat maps. These representations help uncover infrastructure relationships, campaign patterns, and geotemporal dynamics in threat activity. When powered by large language models, MAP enables flexible and context-aware extraction of relational triples from unstructured threat reports.

\begin{tcolorbox}[title=Prompt 7. Threat Knowledge Mapping Prompt]
\textbf{System Prompt:}  
You are a cybersecurity knowledge graph assistant. Extract and relate key entities from the given threat report to form subject-predicate-object triples.

\textbf{Instructions:}  
Identify entities (e.g., threat actors, tools, organizations, IP addresses) and the relationships between them (e.g., "uses", "targets", "associated with").

\textbf{Output:}  
Return a list of triples in the format:  
\texttt{[subject, predicate, object]}  
Include a confidence score (0–1) if applicable.
\end{tcolorbox}

\subsection{RAG (Retrieval-Augmented Generation)}

To enhance generation with accurate and recent data, RAG combines LLM output with real-time retrieval from external threat intelligence APIs or databases. It is particularly useful for describing evolving threats or identifying actor affiliations.

\begin{tcolorbox}[title=Prompt 4. Structured Query for Retrieval]
\textbf{System Prompt:}  
You are a cybersecurity assistant. Formulate a concise search query to retrieve current information about the topic specified below.

\textbf{Instructions:}  
Based on the topic \textit{“Recent activity by APT29 involving phishing attacks”}, generate a query such as:

\textit{“APT29 phishing campaign 2024 indicators, tools, and targets site:mitre.org OR site:virustotal.com”}

\textbf{Output:}  
Return the final query string and optionally list key evidence passages from results.
\end{tcolorbox}

\subsection{SPA (Text Span Localization)}

To precisely extract actionable phrases—such as indicators of compromise or technique descriptions—from long-form cybersecurity text, \textbf{Text Span Localization} (SPA) models are used.

Two key metrics evaluate SPA effectiveness:

\begin{itemize}
  \item \textbf{Exact Match (EM)}:
  \[
  \text{EM} = \frac{\text{Number of exact matches}}{\text{Total predictions}}
  \]
  \item \textbf{Intersection over Union (IoU)}:
  \[
  \text{IoU} = \frac{|S_p \cap S_t|}{|S_p \cup S_t|}
  \]
\end{itemize}

These metrics assess both strict and partial correctness, aiding in accurate downstream processing such as relation extraction or automated summarization.

\begin{tcolorbox}[title=Prompt 5. Span Extraction Prompt]
\textbf{System Prompt:}  
You are a cybersecurity span identification assistant. Extract the text span that describes the primary technique used in the attack.

\textbf{Instructions:}  
Given a report excerpt, locate and return the sentence or phrase that directly describes how the attacker compromised the system (e.g., phishing, lateral movement, privilege escalation).

\textbf{Output:}  
Return the extracted span as plain text.
\end{tcolorbox}

\subsection{CLS (Classification)}

To measure the ability of a system to categorize cybersecurity-relevant textual inputs—such as threat alerts, vulnerability descriptions, or log messages—into predefined classes (e.g., threat categories, severity levels, or attack types), classification models are employed. This is commonly performed using transformer-based large language models (LLMs), which utilize a special token (e.g., [CLS]) to represent sentence-level semantics. The resulting embedding is mapped to labels through a learned classifier.




\subsection{MATH (Mathematical Calculation)}

To perform quantitative analyses and structured computations relevant to cybersecurity, the \textbf{MATH} function supports tasks such as frequency modeling, impact scoring, cryptographic evaluation, and automated threat prioritization. These computations are critical for risk-informed decision-making within cyber threat intelligence pipelines.

A prominent example is the \textbf{Common Vulnerability Scoring System (CVSS v3.1)}, which uses a combination of weighted factors and conditional logic to produce a standardized severity score for vulnerabilities. One key element is the \textit{Base Score}, calculated using the Impact and Exploitability sub scores:

\[
\text{Base Score} =
\begin{cases}
0, & \text{if Impact Subscore } \le 0 \\
\text{RoundUp}\left(\min(\text{Impact} + \text{Exploitability}, 10)\right), & \text{if Scope is Unchanged} \\
\text{RoundUp}\left(\min(1.08 \times (\text{Impact} + \text{Exploitability}), 10)\right), & \text{if Scope is Changed}
\end{cases}
\]

The \textit{Impact Subscore} is computed from confidentiality, integrity, and availability impact metrics as:

\[
\text{ISC}_{\text{Base}} = 1 - (1 - C) \times (1 - I) \times (1 - A)
\]

This formula models the probability that the system's security properties are affected by a vulnerability. The resulting score guides patching priority, risk exposure assessments, and automated vulnerability triage.

Such logic-heavy, non-trivial calculations exemplify the role of mathematical modules in operational cybersecurity settings and justify the integration of computational reasoning capabilities in modern cyber AI systems.

\begin{tcolorbox}[title=Prompt 9. CVSS Score Computation Prompt]
\textbf{System Prompt:}  
You are a cybersecurity scoring assistant. Given a vulnerability description and metric values (Confidentiality, Integrity, Availability, Scope, Attack Vector, etc.), compute the CVSS v3.1 Base Score.

\textbf{Instructions:}  
Use the official CVSS equations and apply the rounding rules specified in the standard. Return both the numeric score and a textual explanation of the computation steps.

\textbf{Output:}  
Return the Base Score as a float (1 decimal place) and a step-by-step explanation.
\end{tcolorbox}

\section{Metric}
\label{app:metric}

\subsection{Sim (BERT Score)}

To evaluate the semantic similarity between cybersecurity-related texts—such as comparing analyst-written threat summaries, aligning generated incident narratives with original reports, or verifying paraphrased explanations of threat indicators—the \textbf{Sim} function utilizes contextual embedding-based metrics. Specifically, it computes \textbf{BERTScore}~\cite{bertscore}, which has been shown to correlate strongly with human judgment in natural language generation tasks.

BERTScore measures semantic equivalence at the token level by aligning contextual embeddings from pre-trained transformer models. The score is computed as:

\[
\text{BERTScore} = \frac{1}{|x|} \sum_{i} \max_j \cos(\mathbf{x_i}, \mathbf{y_j})
\]

where $\mathbf{x_i}$ and $\mathbf{y_j}$ are contextual embeddings of tokens in the candidate and reference texts, respectively. The final score reflects the average of maximal cosine similarities for each token in the candidate sentence.

This metric is particularly valuable in evaluating machine-generated text in cybersecurity domains, where surface-level similarity may fail to capture the deeper equivalence of technical meaning or threat context.

\subsection{Pass (Code Execution Passing Rate)}

To measure the reliability and functional correctness of cybersecurity automation artifacts—such as detection rules, analysis scripts, or integration workflows—the \textbf{Pass Rate} metric is employed. It quantifies how well a system performs under test by evaluating the proportion of test cases that execute successfully within a defined execution cycle, often conducted in a continuous integration (CI) pipeline.

Formally, the Pass Rate is defined as:

\[
\text{Pass Rate} = \frac{\text{Number of Passed Tests}}{\text{Total Tests Executed}} \times 100\%
\]

This metric provides a coarse yet effective indicator of operational readiness. A high Pass Rate implies that the deployed codebase functions as intended across its tested scenarios, which is critical in cybersecurity contexts where automation is used to process threat intelligence, detect anomalies, or trigger incident response mechanisms.

Routine monitoring of this metric supports the early identification of integration regressions, promotes pipeline stability, and ensures confidence in deploying automated defensive measures to production environments.

\subsection{Hit (Top-k Hit Ratio)}

To evaluate the effectiveness of cybersecurity recommendation or retrieval systems—such as those that propose relevant threat indicators, patch suggestions, attack techniques, or investigative leads—the \textbf{Top-k Hit Ratio} is employed. This metric measures how frequently at least one correct or relevant item appears within the top-$k$ ranked results returned by the system.

Mathematically, the Top-k Hit Ratio is defined as:

\[
\text{Hit@}k = \frac{\text{Number of queries with at least one relevant item in top }k}{\text{Total number of queries}}
\]

A higher Hit@k indicates better system performance in surfacing relevant intelligence near the top of recommendations, which is critical for time-sensitive security operations.

\vspace{0.5em}
\noindent\textbf{Use Case Example:}  
If a system recommends threat indicators based on a query about a ransomware family, Hit@5 evaluates whether at least one valid IOC (e.g., file hash or C2 domain) appears in the top 5 returned items.

\begin{tcolorbox}[title=Prompt 6. Hit Evaluation Prompt for Threat Retrieval]
\textbf{System Prompt:}  
You are an assistant for evaluating cybersecurity retrieval systems. Given a query and a list of system-generated recommendations, check whether any ground truth item appears within the top-$k$ returned results.

\textbf{Instructions:}  
For each query, compare the top-$k$ predicted items against the gold-standard set. Indicate \texttt{"Hit"} if at least one match exists, otherwise \texttt{"Miss"}.

\textbf{Output:}  
Return a JSON object with fields:
\texttt{query}, \texttt{top\_k\_results}, \texttt{ground\_truth}, \texttt{hit@k}: true/false
\end{tcolorbox}

\subsection{Dist (Normalized Distance Similarity)}

To evaluate the accuracy of numeric predictions in range-based estimation tasks, such as severity scoring, the \textbf{Normalized Distance Similarity} (\textbf{Dist}) metric is employed. This metric compares the predicted number and the ground-truth and scales the similarity into the $[0, 1]$ range, where higher values indicate closer alignment.

Formally, the similarity is computed as:

$$
\text{Similarity} = 1 - \frac{|\hat{c} - c|}{R}
$$

where $\hat{c}$ and $c$ denote the midpoints of the predicted and true ranges, respectively, and $R$ is the maximum possible value of the range (e.g., 10 in our case of CVSS scores). The metric reflects the Euclidean distance between prediction and truth, normalized such that a perfect match yields a similarity of 1, and the furthest possible discrepancy yields 0.

\section{Additional Experimental Results}
\label{app:expt}

This section presents additional experimental results that complement our main findings, offering deeper insights into model behavior across varied threat-hunting scenarios.

\begin{figure}[!tp]
    \centering
    \includegraphics[width=\textwidth]{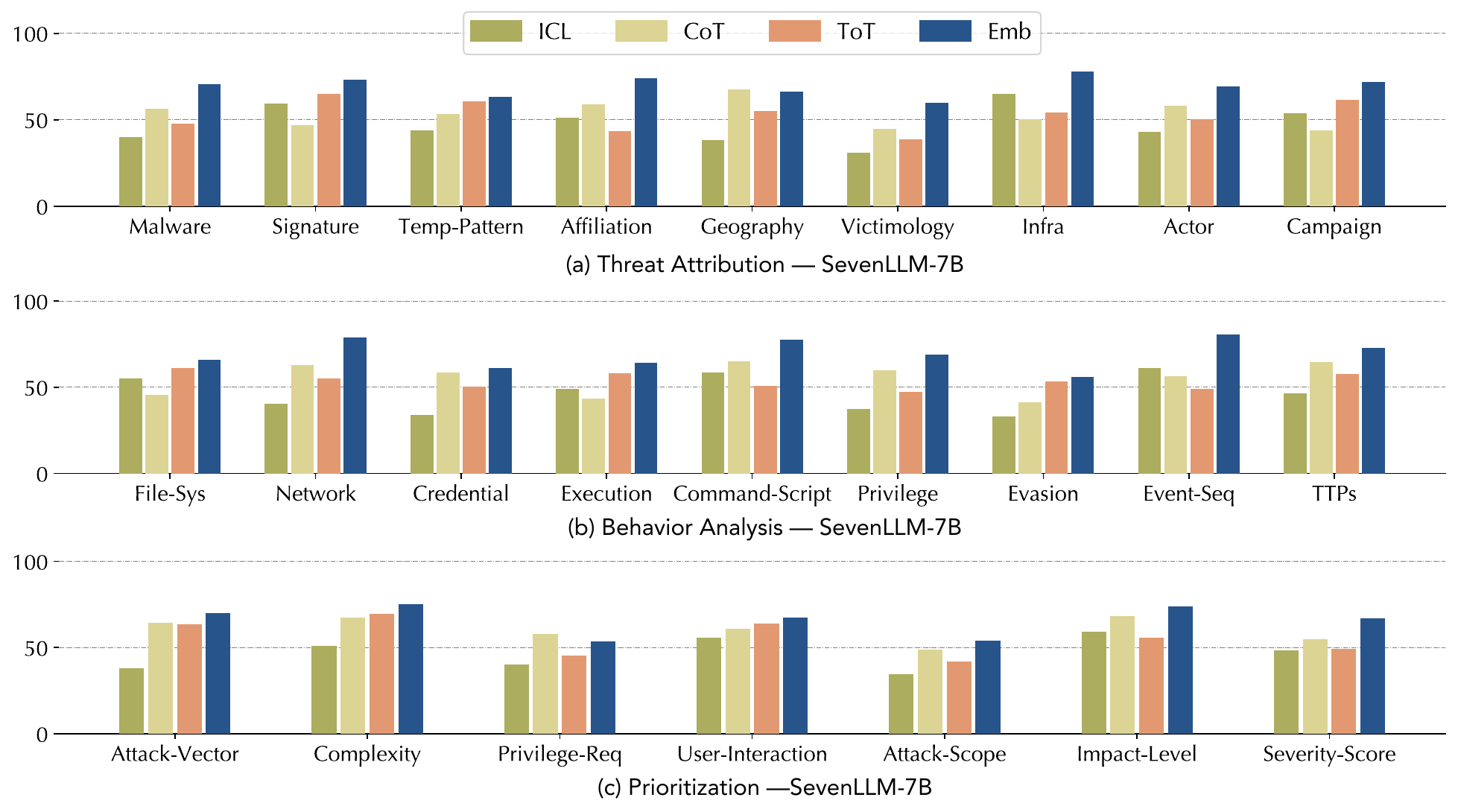}
    \caption{Threat-hunting performance on individual tasks, evaluating under SevenLLM-7B.}
    \label{fig:expt-task-sevenllm}
\end{figure}

\begin{figure}[!tp]
    \centering
    \includegraphics[width=\textwidth]{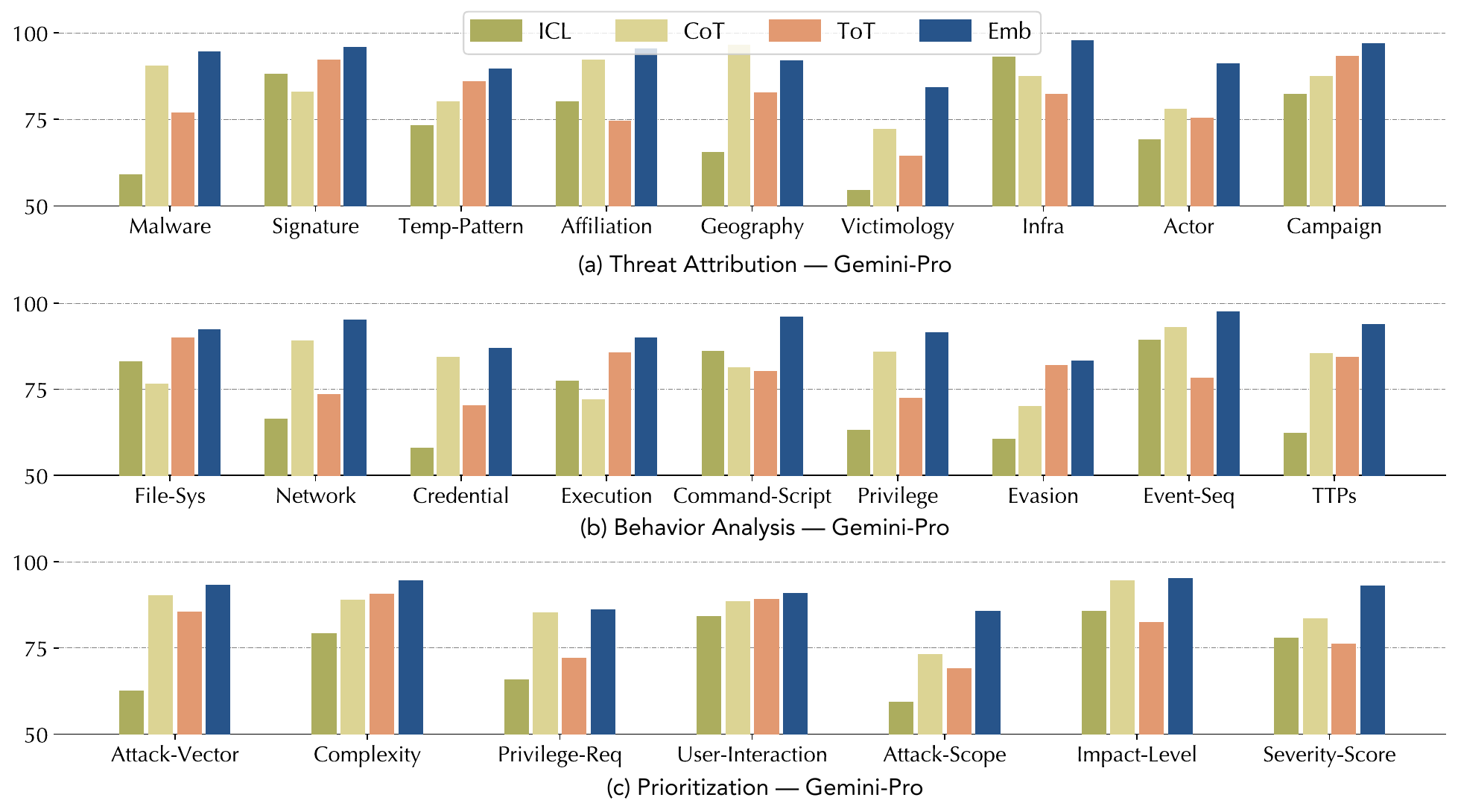}
    \caption{Threat-hunting performance on individual tasks, evaluating under Gemini-pro.}
    \label{fig:expt-task-gemini}
\end{figure}

\begin{figure}[!tp]
    \centering
    \includegraphics[width=\textwidth]{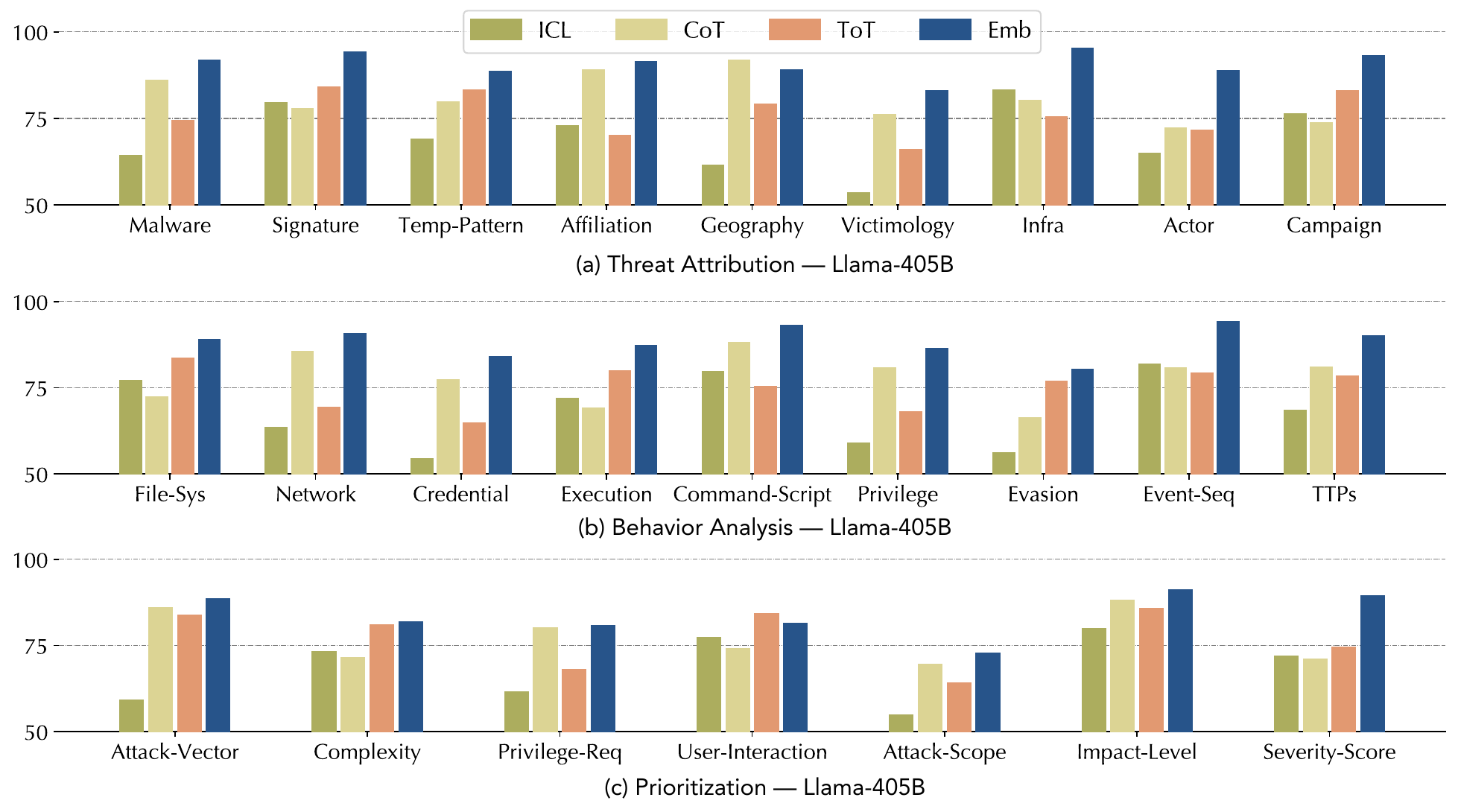}
    \caption{Threat-hunting performance on individual tasks, evaluating under Llama-405B.}
    \label{fig:expt-task-llama}
\end{figure}

{\bf Individual Threat Hunting Performance.} Figure \ref{fig:expt-task-sevenllm}, \ref{fig:expt-task-gemini}, and \ref{fig:expt-task-llama} complement the results as present in Figure \ref{fig:expt-task}, offering aligned insights as exhibited in previous experiments.

\end{document}